\documentclass[preprint,showpacs,preprintnumbers,amsmath,amssymb]{revtex4}

\usepackage{graphicx}
\usepackage{dcolumn}
\usepackage{bm}

\begin{document}

\preprint{APS/123-QED}

\title{One-dimensional particle simulation of the filamentation instability: 
electrostatic field driven by the magnetic pressure gradient force}%
Force line breaks with \\

\author{M. E. Dieckmann}
\affiliation{Centre for Plasma Physics, Queen's University Belfast, Belfast
BT7 1NN, U K
}%

\author{I. Kourakis}
\affiliation{Centre for Plasma Physics, Queen's University Belfast, Belfast
BT7 1NN, U K
}%

\author{M. Borghesi}
\affiliation{Centre for Plasma Physics, Queen's University Belfast, Belfast
BT7 1NN, U K
}%

\author{G. Rowlands}%
 
\affiliation{Physics Department, Warwick University, Coventry CV4 7AL, U K}%

\date{\today}

\begin{abstract}
Two counter-propagating cool and equally dense electron beams are modelled
with particle-in-cell (PIC) simulations. The electron beam filamentation
instability is examined in one spatial dimension, which is an approximation 
for a quasi-planar filament boundary. It is confirmed, that the force on 
the electrons imposed by the electrostatic field, which develops during the 
nonlinear stage of the instability, oscillates around a mean value that 
equals the magnetic pressure gradient force. The forces acting on the 
electrons due to the electrostatic and the magnetic field have a similar 
strength. The electrostatic field reduces the confining force close to the 
stable equilibrium of each filament and increases it farther away, limiting 
the peak density. The confining time-averaged total potential permits an 
overlap of current filaments with an opposite flow direction. 
\end{abstract}

\maketitle

The electron beam filamentation instability (FI) generates magnetic fields 
in energetic astrophysical\textsuperscript{\onlinecite{Gallant,Kazimura}} 
and solar flare plasmas\textsuperscript{\onlinecite{Karlicky}} and in laser 
plasma interactions\textsuperscript{\onlinecite{Tabak,Macchi2}}, if the beam 
speeds $|\mathbf{v}_b|$ are comparable to $c$ and if the densities of the 
counterstreaming beams are similar\textsuperscript{\onlinecite{BreGre}}. It 
has been investigated with one-dimensional (1D) PIC and Vlasov 
simulations\textsuperscript{\onlinecite{Cal2,DavidsonO}} and with 
two-dimensional (2D) PIC simulations\textsuperscript{\onlinecite{Lampe,Pukhov}}.
The counterstreaming electron beam instability has also been examined with a 3D 
PIC simulation\textsuperscript{\onlinecite{Sakai}}. Mobile ions and a guiding 
magnetic field have been taken into 
account\textsuperscript{\onlinecite{Cal2,Pukhov,Stockem}} and statistical 
properties of the FI have been 
obtained\textsuperscript{\onlinecite{Silva,Dieckmann,Rowlands}}.

The FI triggers the growth of waves with the wavevectors $\mathbf{k} \perp 
\mathbf{v}_b$ over a wide band of $k=|\mathbf{k}|$, where the wavenumbers 
$k$ are of the order of the inverse electron skin depth. The electrons are 
deflected by the magnetic field perturbation, and electrons moving in 
opposite directions separate in space. The net current of these flow 
channels amplifies the initial perturbation and, thus, the tendency to 
form current channels. The magnetic field amplitude grows exponentially 
and it saturates by the magnetic trapping of 
electrons\textsuperscript{\onlinecite{DavidsonO}}. The FI can also couple 
nonlinearly to electrostatic 
waves\textsuperscript{\onlinecite{Pukhov,Cal2,Rowlands}}. 
It has been suggested\textsuperscript{\onlinecite{Stockem,Rowlands}} that it 
is the magnetic pressure gradient that gives rise to the electrostatic 
field that grows, when the FI saturates, but it has not yet been
demonstrated quantitatively. This is the purpose of this paper. 

We consider here the FI driven by equally dense and warm electron beams,
which have a Maxwellian velocity distribution in their rest frame. This
case is important, because the growth rate of the FI is highest relative 
to the competing mixed mode and two-stream 
instabilities for symmetric beams\textsuperscript{\onlinecite{BreGre}}. 
We study the FI with a particle-in-cell simulation 
code\textsuperscript{\onlinecite{Dawson}} that is based on the 
electromagnetic and relativistic virtual particle-mesh 
method\textsuperscript{\onlinecite{Eastwood}}.

The FI is modelled in a simulation reference frame, in which both beams 
move into opposite directions at the speed modulus $v_b=0.3c$. 
We isolate the FI by selecting a 1D simulation box that is oriented 
orthogonally to the beam velocity vector $\mathbf{v}_b$ and we resolve
all velocity components. This is an approximation for a quasi-planar 
boundary between filaments with an oppositely directed electron flow. They 
occur in warm plasmas, if the confining magnetic field cannot overcome the 
thermal pressure and they are characterized by planar magnetic 
fields\textsuperscript{\onlinecite{Stockem,Dieckmann,SilvaAIP}}. The 
periodic boundary conditions of the short simulation box result in the 
development of only one pair of filaments. The restriction to one dimension 
inhibits the merging of the filaments\textsuperscript{\onlinecite{Lampe}} 
and we can analyse the relation between the electric and magnetic fields 
of the quasi-stationary filaments.

Beam 1 has the mean speed $\bm{v}_{b1} = v_b \bm{z}$ and the beam 2 
has $\bm{v}_{b2} = -v_b \bm{z}$. Both beams are spatially uniform 
and have a Maxwellian velocity distribution in their respective rest frame 
with a thermal speed $v_{th}={(k_b T / m_e)}^{0.5}$ of $v_b / v_{th} = 18$.  
The 1D simulation box with its periodic boundary conditions is aligned 
with the $\bm{x}$-direction. We thus denote positions by the scalar 
$x$. The plasma frequency of each beam with the number density $n_e$ is 
$\omega_p = {(e^2 n_e / m_e \epsilon_0)}^{0.5}$. The total plasma frequency 
$\Omega_p =\sqrt{2}\omega_p$. The electric and magnetic fields are
normalized to $\bm{E}_N = e\bm{E}/c m_e \Omega_p$ and $\bm{B}_N 
= e\bm{B} /m_e \Omega_p$ and the current to $\bm{J}_N = \bm{J} 
/ 2 n_e e c$. The physical position and time are normalized as $x_N = x/ 
\lambda_s$ with the electron skin depth $\lambda_s = c / \Omega_p$ and 
$t_N = t \Omega_p$. We drop the indices $N$ and $x,t,\bm{E},\bm{B},
\bm{J}$ are specified in normalized units.

The box length $L=0.89$ is resolved by $N_g = 500$ grid cells with the 
length $\Delta_x$. The simulation time $t_S = 125$. The phase space 
distributions $f_1(x,\bm{p})$ of beam 1 and $f_2(x,\bm{p})$ of beam 2 
are each sampled by $N_p = 6.05 \cdot 10^7$ computational particles 
(CPs). The total phase space density is defined as $f(x,\bm{p})=
f_1(x,\bm{p})+ f_2(x,\bm{p})$.

\begin{figure}
\begin{center}
\includegraphics[width=3in]{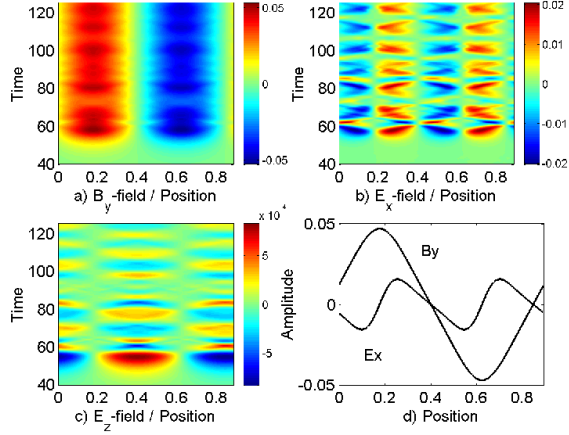}
\caption{(Color online) (a)-(c) show $B_y$, $E_x$ and $E_z$. All fields 
are stationary in space and $B_y$ is quasi-stationary also in time for 
$t>56$. $E_x$ oscillates in space twice as fast as $B_y$ and both are 
spatially correlated for $55<t<125$. The phase of $E_z$ is shifted by 
$90^\circ$ relative to that of $B_y$ and it is damped. The $B_y,E_x$ fields 
at $t=56$ are compared in (d).
\label{fig4}}
\end{center}
\end{figure}

The electrons and their micro-currents are redistributed by the FI along 
$x$. The charge- and current-neutral plasma is transformed into one with 
$J_z (x) \neq 0$. The z-component of Ampere's law is in the 1D geometry 
$\partial_x B_y = J_z + \partial_t E_z$. A $J_z \propto \sin{(kx)}$ gives 
a $E_z \propto \sin{(kx)}$ and $B_y \propto -\cos{(kx)}$ so that $E_z$ 
and $B_y$ will have a phase shift of $90^\circ$. Figure \ref{fig4} reveals 
this phase shift between $B_y$ and the evanescent $E_z$. It also shows, 
that an electrostatic $E_x$-field grows. The $B_y (x,t)$ and the $E_x (x,t)$ 
oscillate in space with the wavenumbers $k_1$ and $k_2$, respectively, where 
$k_j = 2\pi j /L$. Both fields are spatially correlated. The comparison of 
$E_x$ and $B_y$ at $t=56$ demonstrates that $E_x = 0$ if $B_y=0$ or if 
$d_x B_y = 0$.

We determine now the relation between $E_x$ and $B_y$. Let $E_B (x,t)$ 
be an electric field along $x$, which excerts the same force on an electron 
as the magnetic pressure gradient force does. This electric field is given 
in our normalization (charge  $q=-1$) as $E_B (x,t) = -B_y(x,t) \, d_x 
B_y(x,t)$. We note that $B_y(x,t>56)$ is 
quasi-stationary, while $E_x (x,t>56)$ oscillates in time. The oscillation 
amplitude of $E_x (x,t>56)$ is approximately constant and it apparently 
oscillates around a stationary background field. It is helpful to average 
the $E_x (x,t)$ and the $E_B(x,t)$ over the time interval $t_1=56$ to $t_2 
= 125$ to give $\tilde{E}_x (x) = {(t_2-t_1)}^{-1} \int_{t_1}^{t_2} E_x (x,t) 
\, dt$ and $\tilde{E}_B (x) = {(t_2-t_1)}^{-1} \int_{t_1}^{t_2} E_B (x,t) \, dt$. 

\begin{figure}
\begin{center}
\includegraphics[width=3in]{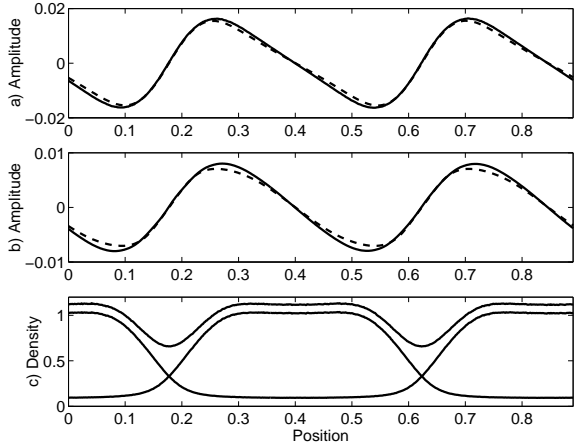}
\caption{(a) The $E_x (x,t=56)$ (dashed curve) and $2E_B(x,t=56)$ 
(solid curve). (b) The time-averaged $\tilde{E}_x$ (dashed curve) and 
$\tilde{E}_B$ (solid curve). (c) The number densities for $t=56$, 
normalized to $2n_e$, of both beams separately (beam 1 is almost 
confined to $0.2 < x < 0.6$) and both densities added together.\label{fig6}}
\end{center}
\end{figure}

Figure \ref{fig6}(a) displays the $E_x (x,t=56)$ when the FI has just 
saturated and reached its peak amplitude and it compares it with 
$E_B(x,t=56)$. 
It turns out that $E_x (x,t=56) \approx 2E_B (x,t=56)$. The time-averaged
fields fulfill $\tilde{E}_x (x) \approx \tilde{E}_B(x)$ in Fig. \ref{fig6}(b). 
The $E_x(x,t>56)$ oscillates in time with an amplitude $\approx \tilde{E}_B(x)$ 
around a stationary background field with the amplitude $\approx \tilde{E}_B
(x)$. Both amplitudes add up to $2\tilde{E}_B(x)$ at $t=56$. When the 
oscillatory and the background electric field have a phase shift of 
$180^\circ$ in time, they result in a $E_x (x,t_c) \approx 0$, for example 
when $t_c=75$ in Fig. \ref{fig4}(b). 

The $E_B(x,t=56)$ and $\tilde{E}_B (x)$ 
correlate well in Fig. \ref{fig6}(c) with the normalized number density 
distributions $n_{1,2}(x) = {(2n_e)}^{-1} \int f_{1,2} (x,\bm{p})\, d\bm{p}$ 
of each beam and also with the summed distribution $n_1(x)+n_2(x)$ at $t=56$. 
The total density is modulated by about $30\%$, while that of $n_1$ and 
$n_2$ varies by an order of magnitude.

\begin{figure}
\begin{center}
\includegraphics[width=3in]{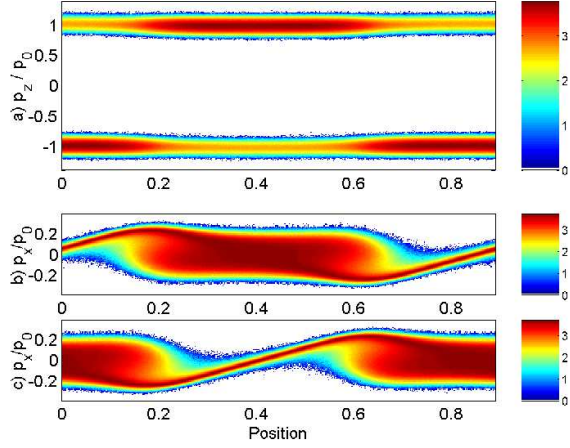}
\caption{(Color online) The 10-logarithmic phase space densities in units 
of CPs at $t=56$: Panel (a) shows the $f(x,p_z)$ with $p_0 = m_e v_b \Gamma 
(v_b)$. The temperature and the mean velocity along $z$ of the electrons 
are unchanged as a function of $x$. The density oscillates by 
the factor $\approx 10$. The $f_1(x,p_x)$ is shown in (b) and the $f_2(x,p_x)$ 
in (c).\label{fig7}}
\end{center}
\end{figure}

Figure \ref{fig7} shows the electron phase space distributions $f(x,p_z)$, 
$f_1(x,p_x)$ and $f_2(x,p_x)$ at the time $t=56$.
The mean velocity along $z$ of the electrons of the beams 1 (i=1) and 2 
(i=2) is practically constant as a function of $x$. Any spatial modulation 
would be caused by the $\bm{E} \times \bm{B}$-force, which is given by the 
product of $E_x$ and $B_y$ in our geometry. The effects of this force are
small.

The supplementary movie\textsuperscript{\onlinecite{Movie}} animates in 
time the evolution of $f_1(x,p_x)$ and $f_1(x,p_z)$, where the color scale 
denotes the 10-logarithmic number of CPs. The electrons are redistributed 
along $x$ but they keep their $p_z$ almost unchanged. Their flow along $x$ 
oscillates, giving a $J_x (x,t)\neq 0$. The $f_1(x,p_x)$ has a dense 
electron core, which rotates in the $x,p_x$-plane around $x=0.4$. Two phase 
space vortices are convected with this rotating flow. The phase space motion 
of the electrons around the equilibrium points $x_e$ with $E_x(x_e) = 
B_y(x_e) = 0$, for example $x_e=0.4$ in the supplementary movie, reveals, 
that they are trapped by a potential. 

We can estimate the contributions of $E_x$ and $B_y$ to this potential 
after the saturation of the FI. We average the fields $\tilde{E}_x (x) 
= {(t_2-t_1)}^{-1} \int_{t_1}^{t_2} E_x (x,t) \, dt$ and $\tilde{B}_y(x) = 
{(t_2-t_1)}^{-1} \int_{t_1}^{t_2} B_y (x,t) \, dt$ in time from $t_1 = 68$ to 
$t_2 = 125$. In what follows we consider beam 1 with $v_b>0$. According 
to the supplementary movie, most of the electrons have the velocity 
components $v_z \approx v_b$ and $v_x \ll v_b$. The electrons retain their 
initial $v_y \ll v_b$, since no force component along $y$ develops. The 
dominant component of the time-averaged magnetic force is thus $\tilde{F}_x 
= v_b \tilde{B}_y$ for $q=-1$. The time-averaged force along $x$ 
is then $\tilde{F}_x = -(\tilde{E}_{x} +\tilde{E}_D)$ with $\tilde{E}_D = 
-v_b \tilde{B}_y$ and we define $\tilde{E}_T = \tilde{E}_x + \tilde{E}_D$. 
The time-averaged potentials $\tilde{U}_j(x) = U_{0,j}
+\int_0^x \tilde{E}_j (\tilde{x}) d\tilde{x}$ with the indices $j=x,D,T$ 
are calculated from these fields and $U_{0,j}$ is set such that 
$\tilde{U}_j(x_e=0.4)=0$. The potentials are given in Volts.

\begin{figure}
\begin{center}
\includegraphics[width=3in]{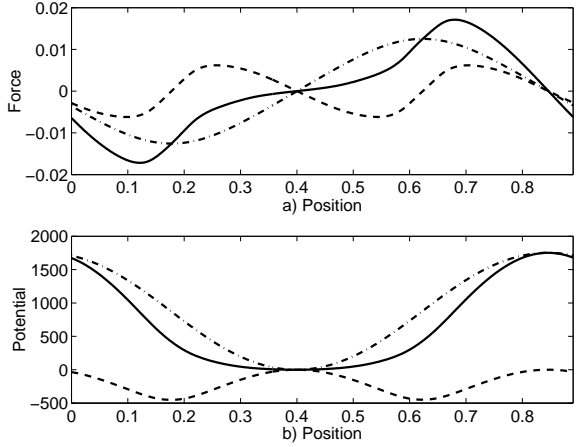}
\caption{The fields $\tilde{E}_j$ and the potentials $\tilde{U}_j$ 
averaged over $68<t<125$: (a) shows $\tilde{E}_x$ (dashed), $\tilde{E}_D 
= -v_b \tilde{B}_y$ (dash-dotted) and $\tilde{E}_T = \tilde{E}_x+\tilde{E}_D$ 
(solid line). Positive 
$\tilde{E}_j$ accelerate electrons into the negative $x$-direction. (b) 
shows the potential $\tilde{U}_x$ (dashed), $\tilde{U}_D$ (dash-dotted) 
and the $\tilde{U}_T$ (solid). The potential at $x=0.4$ is the reference 
potential.\label{fig8}} 
\end{center}
\end{figure}

Figure \ref{fig8} displays the time-averaged fields and potentials. 
The $\tilde{E}_x$ destabilizes the equilibrium position $x_e = 0.4$, 
because the negative $\tilde{E}_x (x>x_e)$ close to $x_e$ accelerates the 
electron in the positive direction and the positive $\tilde{E}_x (x<x_e)$ 
close to $x_e$ in the negative direction. The $\tilde{E}_D (x\approx x_e)$ 
is confining the electrons around $x\approx x_e$. The $|\tilde{E}_D|>
|\tilde{E}_x|$ for $x\approx x_e$ and $E_T$ is thus a confining force. 
However, the electron acceleration at $x \approx x_e$ is decreased by 
$\tilde{E}_x$ and increased at larger $|x-x_e|$. 

The CPs of the beam 1 should follow almost straight paths close to $x_e=0.4$ 
and they should be rapidly reflected for $|x-x_e|>0.2$. The potential 
difference $\Delta_U = max(\tilde{U}_T)-min(\tilde{U}_T) \approx$ 1700 V 
should trap electrons with speeds up to $\Delta_v ={(2 e \Delta_U 
/ m_e)}^{1/2} / v_b \approx 0.27$. This matches the momentum spread of the 
cool core population in Fig. \ref{fig7} and in the supplementary movie. 
The oscillations of $E_x$ in Fig. \ref{fig4} and, thus, of the strength of
the confining potential explain the periodic release of electrons from 
this cool core seen in the supplementary movie. The oscillatory force 
imposed on the electrons by $E_x(x,t)$ contributes to their heating.

\begin{figure}
\begin{center}
\includegraphics[width=3in]{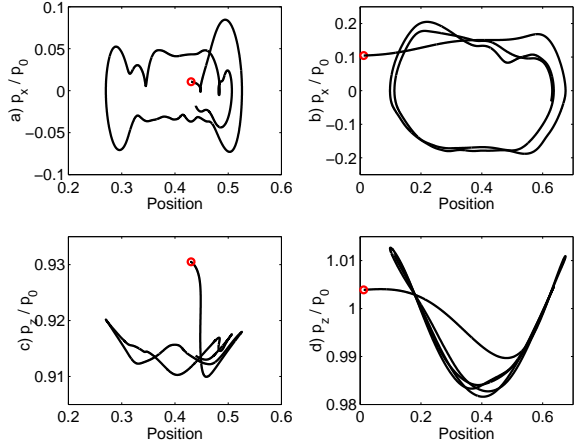}
\caption{The trajectories of two selected CPs: (a),(c) show the
$x,p_x$ and $x,p_z$ diagrams of the CP 1. (b),(d) show the corresponding 
diagrams for the CP 2. The circle denotes the starting point of the 
trajectory. Both CPs follow straight paths in the $x,p_x$ plane for
$0.33<x<0.47$ and they are rapidly reflected outside this interval.
\label{fig9}}
\end{center}
\end{figure}

Figure \ref{fig9} follows the trajectories of two CPs of beam 1.
The circles denote the times, when the CPs start interacting with the 
fields and the trajectories are followed until $t=125$. The CP 1 has a 
low initial modulus of $p_x$ and CP 2 a high one. Both CPs follow straight 
paths in the interval $0.33 < x < 0.47$, in which $\tilde{E}_T$ in Fig.
\ref{fig8} is small. The phase space path of the faster CP 2 is smoother
than that of CP 1. The low speed of CP 1 implies a long crossing time of 
the interval with a low modulus of $\tilde{E}_T$ and the CP 1 experiences 
several oscillation cycles of $E_x$. Both CPs are reflected outside the 
interval $0.33 < x < 0.47$ and they remain trapped, because they can not 
overcome the potential difference $\Delta_U$. Both electrons change their
$v_z$ only by a few percent and $|v_x| < v_b / 5$, which is supporting 
our previous assumption of a dominant and constant force $v_b \tilde{B}_y$ 
along $x$.

In summary, we have examined the saturation of the filamentation 
instability (FI) driven by two counter-propagating, weakly relativistic 
and symmetric beams of electrons. The 1D simulation box has been oriented 
orthogonally to $\bm{v}_b$. It can approximate the quasi-planar boundary 
between two filaments with oppositely directed flow, which shows up if 
the magnetic confinement cannot overcome the thermal pressure. This 
geometry is beneficial, because two of the three spatial derivatives in 
the Maxwell equations vanish, which separates the electrostatic and the 
electromagnetic fields. 

We have confirmed here, that the electrostatic field, which grows during 
the nonlinear phase of the FI and in a 1D geometry, is driven by the 
magnetic pressure gradient. This has been proposed 
elsewhere\textsuperscript{\onlinecite{Rowlands,Stockem}}, but a quantitative 
comparison has so far been lacking. We have shown with a PIC simulation 
that the force imposed on an electron by the time-averaged electrostatic 
field $\tilde{E}_x (x)$ matches the $\tilde{E}_B(x)$, which results from 
the time-averaged magnetic pressure gradient force. The $E_x(x,t)$ is, 
however, not time-stationary, which can be explained as follows. 

The FI accelerates through the magnetic pressure gradient force the 
electrons and a current $J_x(x,t)$ builds up. This current results in 
the 1D geometry with $\nabla \times \bm{B}=0$ through the normalized 
equation $J_x (x,t)= - \partial_t E_x(x,t)$ in a growing $E_x(x,t)$. 
The initial conditions are $E_x (x,t=0)=0$ and $J_x (x,t=0)=0$. Any 
oscillatory solution for $J_x(x,t)$ and $E_x (x,t)$ implies through 
$J_x (x,t)= - \partial_t E_x(x,t)$ that $J_x$ and $E_x$ cannot 
simultaneously oscillate in time around their initial values. The 
$E_x(x,t)$ oscillates instead around its time-average, which is the 
background field $\tilde{E}_B(x)$. The oscillation amplitude of $E_x$ 
is approximately $\tilde{E}_B(x)$. The superposed oscillatory and background 
field thus oscillates between $E_x(x,t_c) = 0$ at certain times $t_c$, 
fulfilling the initial condition at $t_c=0$, and a maximum 
$E_x(x,t_c) = 2\tilde{E}_B(x)$.

We have confirmed previous suggestions, that the electric field force
is comparable to the magnetic field 
force\textsuperscript{\onlinecite{Cal2,Pukhov}}. We have used the
time-averaged electric and magnetic forces to estimate their effects
quantitatively. The electric field repels electrons at the filament 
centres and attracts them if they are farther away, which permits 
filaments to overlap and limits their peak density.

{\bf Acknowledgments:}
The authors acknowledge the financial support by an EPSRC Science 
and Innovation award, by the visiting scientist programme of the
Queen's University Belfast, by VR and by the 
DFG (Forschergruppe FOR1048). The HPC2N 
computer center has provided the computer time.


\end{document}